\documentclass{bmvc2k}
\usepackage{amsthm,amsmath,amssymb}
\usepackage{mathrsfs}
\usepackage{multirow}
\usepackage{marvosym}

\newcommand{\ie}{\textit{i}.\textit{e}. }

\title{Approaching the Limit of Image Rescaling via Flow Guidance}

\addauthor{Shang Li}{lishang2018@ia.ac.cn}{12}
\addauthor{Guixuan Zhang}{guixuan.zhang@ia.ac.cn}{2}
\addauthor{Zhengxiong Luo}{zhengxiong.luo@cripac.ia.ac.cn}{12}
\addauthor{Jie Liu\textsuperscript{\Letter}}{jie.liu@ia.ac.cn}{2} 
\addauthor{Zhi Zeng}{zhi.zeng@ia.ac.cn}{2}
\addauthor{Shuwu Zhang}{shuwu.zhang@ia.ac.cn}{23}

\addinstitution{
	School of Artificial Intelligence\\
	University of Chinese Academy of Sciences\\
	Beijing, China
}
\addinstitution{
	Institute of Automation\\
	Chinese Academy of Sciences\\
	Beijing, China
}
\addinstitution{
	Beijing Engineering Research Center for Digital Content Technology\\
	Beijing, China
}

\runninghead{Li ET AL}{Approaching the Limit of Image Rescaling via Flow Guidance}

\def\eg{\emph{e.g}\bmvaOneDot}

\begin{document}
	
	\maketitle
	
	\begin{abstract}
		Image downscaling and upscaling are two basic rescaling operations. Once the image is downscaled, it is difficult to be reconstructed via upscaling due to the loss of information. To make these two processes more compatible and improve the reconstruction performance, some efforts model them as a joint encoding-decoding task, with the constraint that the downscaled (\ie encoded) low-resolution (LR) image must preserve the original visual appearance. To implement this constraint, most methods guide the downscaling module by supervising it with the bicubically downscaled LR version of the original high-resolution (HR) image. However, this bicubic LR guidance may be suboptimal for the subsequent upscaling (\ie decoding) and restrict the final reconstruction performance. In this paper, instead of directly applying the LR guidance, we propose an additional invertible flow guidance module (FGM), which can transform the downscaled representation to the visually plausible image during downscaling and transform it back during upscaling. Benefiting from the invertibility of FGM, the downscaled representation could get rid of the LR guidance and would not disturb the downscaling-upscaling process. It allows us to remove the restrictions on the downscaling module and optimize the downscaling and upscaling modules in an end-to-end manner. In this way, these two modules could cooperate to maximize the HR reconstruction performance. Extensive experiments demonstrate that the proposed method can achieve state-of-the-art (SotA) performance on both downscaled and reconstructed images.
	\end{abstract}
	
	\section{Introduction} \label{sec:intro}
	With the tremendous advances of mobile devices and web applications, the demand has surged for image downscaling to downscale high-resolution (HR) images to smaller-sized ones, which can save storage, accelerate transmission speed, or fit low-resolution (LR) screens while preserving the original visual appearance.  Meanwhile, the inverse process, image upscaling, is also indispensable to reconstruct the underlying HR image of the downscaled LR input. In previous researches, image downscaling and upscaling are studied as two independent tasks, and many methods have been proposed for downscaling~\cite{bicubic, kopf, og, hou} and upscaling. One of the most well-known downscaling methods is bicubic interpolation~\cite{bicubic}, and upscaling is studied as super-resolution (SR)~~\cite{glasner, srcnn, accuratesr, srgan, nonpara, dan}. However, due to the information loss during downscaling, the reconstruction performance of the subsequent upscaling is heavily restricted~\cite{nyquist}. More and more efforts focus on the impact of downscaling for the final reconstruction performance and attempt to find the upscaling-optimal downscaling method.
	
	To make downscaling and upscaling more compatible, some recent methods model these two processes as a joint encoding-decoding task~\cite{task,cnncr,resampler}. Specifically, the downscaling module is regarded as the encoder to downscale the original HR image into a smaller-sized representation, while the upscaling module is the decoder to reconstruct a more detailed HR image from the downscaled representation. However, there is an additional constraint: the downscaled representation must preserve the visual appearance of the original HR image, that is, it should be a visually plausible downscaled image. To implement this constraint, most existing methods employ the bicubically downscaled version of the original HR image to supervise downscaling module, which is termed bicubic LR guidance~\cite{task}. This guidance helps the downscaling module produce a visually plausible LR image, but this may not be optimal for the subsequent upscaling and restrict the HR reconstruction performance.

	To address this problem, we design a flow guidance rescaling network (FGRN), which can remove the aforementioned restriction on the downscaled representation. Specifically, FGRN is an encoding-decoding based network with an invertible flow guidance module (FGM). FGM can transform the downscaled representation into a visually pleasing LR image during downscaling and transform this image back to the original representation during upscaling, which could guarantee the visual quality of the downscaled image. Moreover, the lossless transformation of FGM could help relax restrictions on the downscaled representation. Thus, the downscaling and upscaling modules can be jointly optimized in an end-to-end manner without applying any LR guidance to the downscaling module. In this way, the downscaled representation could be upscaling-optimal and maximize the HR reconstruction performance.

	In summary, our main contributions are as follows:
	\begin{itemize}
		\item
		To the best of our knowledge, our proposed flow guidance rescaling network (FGRN) is the first image rescaling method without any constraints on the downscaling process. With the help of the invertible flow guidance module, the downscaling and upscaling modules can be optimized in an end-to-end manner and cooperate to maximize the HR reconstruction performance.
		\vspace{-0.02\linewidth}
		\item 
		We design an invertible flow guidance module (FGM), which can perform lossless transformation between downscaled representations and visually favorable LR images. It enables us to remove the restriction on the downscaling process and helps improve the final reconstruction performance.
		\vspace{-0.02\linewidth}
		\item
		Extensive experiments on different benchmark datasets demonstrate that the proposed method can achieve SotA performance on both downscaled and reconstructed images with fewer parameters.
		
	\end{itemize}
	
	\vspace{-0.01\linewidth}
	\section{Related Work}
	\subsection{Image Rescaling}
	Image downscaling and upscaling are two basic image rescaling tasks. Image downscaling aims to reduce the resolution of the HR image while keeping its visual appearance. Image upscaling aims to reconstruct a visually pleasing HR image through the given LR image. In early researches, they are treated as independent tasks. One of the most well-known downscaling methods is Bicubic~\cite{bicubic}. And upscaling is studied as super-resolution (SR)~\cite{glasner} and plays a significant role in low-level vision tasks~\cite{ facesr, reidsr}. However, due to the information loss in downscaling, the performance of the subsequent upscaling is heavily restricted.
	
	Toward this issue, some recent efforts attempt to model these two processes as a united task, thus the downscaling method could be ``customized'' for the upscaling method. Although these methods make downscaling and upscaling cooperate better, there are still some limitations. For instance, \cite{task} and \cite{cnncr}  jointly optimize downscaling and upscaling with an encoding-decoding based framework, and they apply the bicubic LR guidance to guarantee the visual quality of the downscaled image. However, a visually plausible downscaled image may not be the optimal representation for upscaling. Besides, \cite{irn} uses an invertible network to model downscaling and upscaling according to their reciprocal nature. But the invertibility of~\cite{irn} restricts that these two processes must share the same network parameters in opposite directions, which is rigid and may largely limit the representation capability. In this paper, the structure of our rescaling module is encoding-decoding based and free to be adjusted. Moreover, we remove the restriction on the downscaling module by proposing an invertible flow guidance module. It can guarantee the visual quality of the downscaled image and help the rescaling module maximize the HR reconstruction performance. 
	
	\textbf{Difference to SR.} Note that image upscaling in this paper is different from SR. In our scenario, the ground-truth HR image is available at the beginning but somehow is discarded and we store/transmit the LR version instead. Our goal is to recover the HR image using this LR image. While for SR, the real HR is unavailable in applications and the task is to generate new HR images for LR ones.
	
	\vspace{-2mm}
	\subsection{Normalizing Flow}
	Normalizing flows aim to map the complex data distribution $p_X(x)$ to a tractable prior distribution $p_Z(z)$ using the bijective function $F$, that is $z = F(x)$. Thus the generative process over $X$ is defined as $x = F^{-1}(z)$. Given such bijective function $F$, the \textit{change of variable formula} defines the model distribution on $X$ by:
	\begin{equation}\label{eq:1}
		\small
		p_X(x) = p_Z(F(x))\vert \mathrm{det}(\frac{\partial F(x)}{\partial x^T}) \vert
	\end{equation}
	where $\vert \mathrm{det}(\frac{\partial F(x)}{\partial x}) \vert$ is the absolute value of the determinant of the Jacobian matrix $\frac{\partial F(x)}{\partial x}$. A stack of such invertible transformations can form a complex final density. NICE~\cite{nice} is a flow model that stacks non-linear additive coupling and other transformation layers. Based on NICE, RealNVP~\cite{nvp} further upgrades additive coupling to affine coupling without loss of invertibility and achieves better performance. Then Glow~\cite{glow} replace the fixed permutation layer in~\cite{nvp} with the more flexible $1\times 1$ convolutional layer. Nowadays, normalizing flows have been successfully applied in generating images~\cite{flowgan, liuconditional}, audio data~\cite{flowavenet} and point cloud data~\cite{cflow,pointflow}. As for image rescaling, IRN \cite{irn}  is an invertible rescaling network to model image downscaling and upscaling. However, this invertible model only allows these two processes to share the same parameters in opposite directions, and the prior distribution in IRN could not compensate for the information lost in downscaling. In this paper, we use flow to perform lossless transformation just between downscaled representations and visually pleasing LR images, instead of the overall image rescaling network.

	\vspace{-2mm}
	\section{Method}
	\subsection{Bicubic LR guidance}~\label{sec:formuation}
	To alleviate the loss of information during downscaling, recent researches~\cite{irn,task, cnncr} adopts an encoding-decoding based framework. The downscaling module $G_d$ is jointly optimized with the upscaling module $G_u$. They hope that $G_d$ could learn to be upscaling-optimal. The overall downscaling-upscaling process can be expressed as:
	\begin{equation}\label{eq:2}
		\small
		\hat{x} = G_d(y), \qquad \hat{y} = G_u(\hat{x}),
	\end{equation}
	where $y$ is the original HR image, $\hat{x}$ is the downscaled image, and $\hat{y}$ is the reconstructed image. To ensure the reconstruction performance, $\hat{y}$ is directly supervised with $y$ via L1 loss. To ensure that $\hat{x}$ keeps the visual appearance of  $y$,  it is usually supervised with the bicubically downscaled version of $y$, which is what we call bicubic LR guidance. Although $G_d$ is jointly optimized with $G_u$ in these methods,  the reconstruction performance may still be restricted, because the bicubic  LR guidance may get conflict with the reconstruction quality. To maximize the reconstruction performance, it is better to minimize the influence of constraint on $G_d$.

	\begin{figure}[h]
		\centering
		\includegraphics[width=\linewidth]{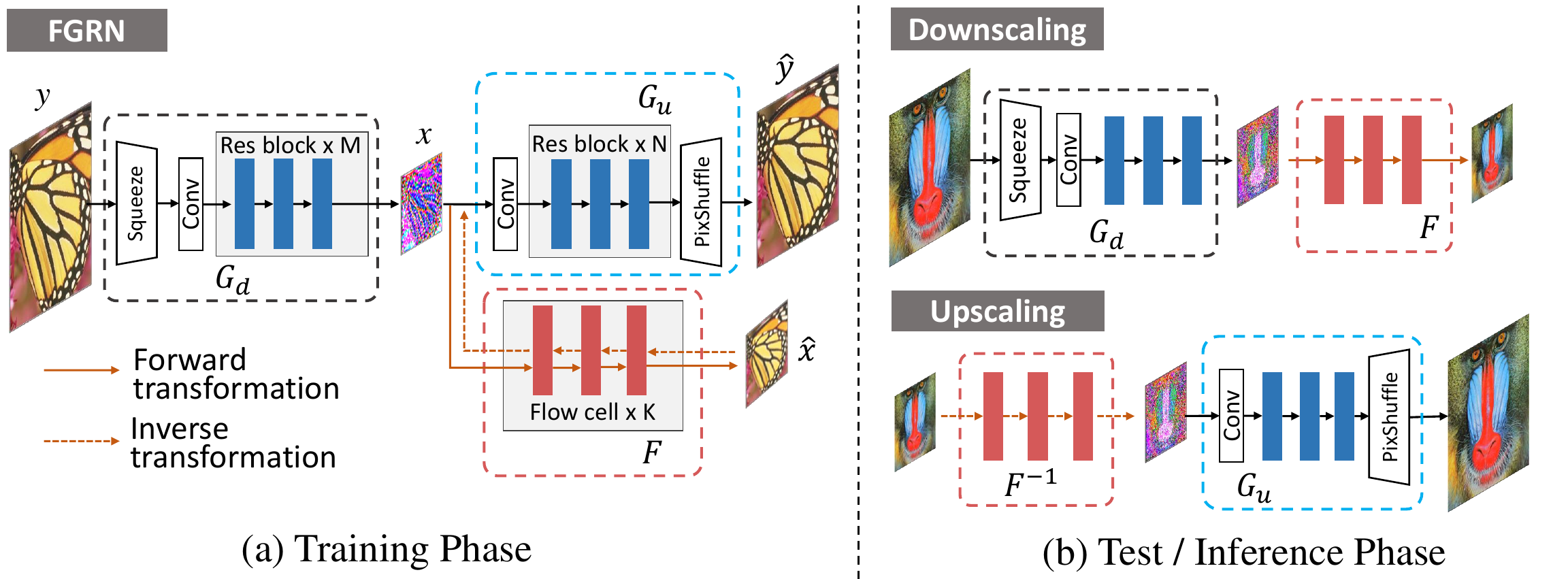}
		\vspace{-0.03\linewidth}
		\caption{The network architecture of FGRN.}\label{fig:FGRN}
		\vspace{-0.03\linewidth}
	\end{figure}
	
	\subsection{Invertible Flow Guidance}
	Different from the methods above, we do not directly apply bicubic LR guidance on $G_d$.  In our method, $G_d$ and $G_u$ are optimized in an end-to-end manner. They are only responsible for improving the quality of the upscaled HR image $\hat{y}$ without considering the visual quality of the downscaled representation, which can be formulated as: 
	\begin{equation}
		\small
		x = G_d(y), \qquad \hat{y} = G_u(x),
	\end{equation}
	where $x$ is downscaled representations. We call it ``representations'' instead of ``image'', because $x$ may not have any visual appearances, since we do not apply any constraint on it. In this way,  it is easier for $G_d$ to become optimal for the upscaling task and help maximize the HR reconstruction performance. It will be experimentally proved in Sec~\ref{sec:diff_loss}.
	
	As for achieving a visually plausible downscaled image, we further introduce an invertible flow guidance which can perform lossless transformation between the downscaled representation and the visually plausible LR image:
	\begin{equation}
		\small
		\begin{split}
			\hat{x} = {F}(x),\quad
			x = F^{-1}(\hat{x}). 
		\end{split}
	\end{equation}
	where $F$ is our flow guidance module (FGM). During downscaling, $F$ can transform the downscaled representation $x$ (\ie $G_d(y)$) to a visually plausible LR image $\hat{x}$. While during upscaling, $F^{-1}$ can inversely transform the given LR image $\hat{x}$ to the original representation $x$ which will be subsequently upscaled into an excellent reconstructed HR image. 
	
	\begin{figure}[t]
		\centering
		\includegraphics[width=0.8\linewidth]{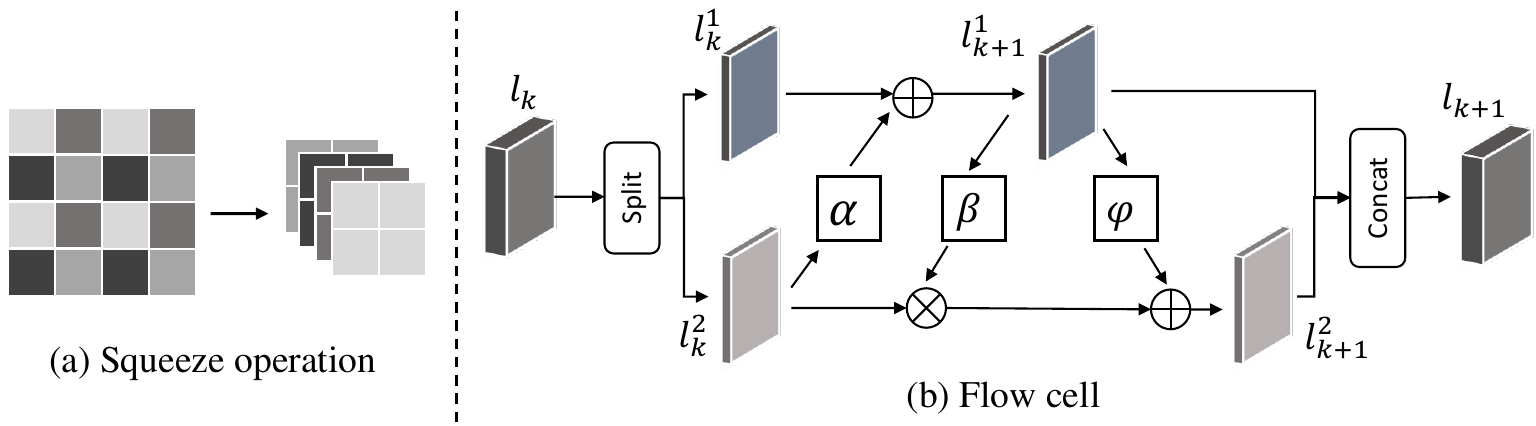}
		\vspace{-0.02\linewidth}
		\caption{(a) The squeeze operation in $G_d$. (b) The structure of flow cell in $F$.}
		\label{fig:FGM}
		\vspace{-0.03\linewidth}
	\end{figure}
	
	\subsection{Network Architecture}
	As can be seen in Figure~\ref{fig:FGRN} (a), the proposed FGRN includes three parts: 1) the downscaling module $G_d$, 2) the upscaling module $G_u$, and 3) the flow guidance module $F$.
	
	\textbf{Downscaling module.} $G_d$ is responsible for mapping the input HR image to a downscaled representation. It consists of a squeeze layer \cite{nvp}, a shallow convolutional layer, and a stack of simple residual blocks (ResBlock)~\cite{edsr,rdn}. As shown in Figure~\ref{fig:FGM} (a), the squeeze layer can reshape the input HR image of size $sH\times sW\times C$ into the representation of size $H\times W\times s^2C$ ($s$ denotes the scale factor), which compresses the spatial resolution but will not cause information loss. 
	
	\textbf{Upscaling module.} $G_u$ aims to reconstruct the original HR image through the downscaled LR image. The structure of $G_u$ is similar to many SR models~\cite{edsr, rcan} that contain a stack of residual blocks for different feature extraction and the pixel shuffle (PixShuffle)~\cite{pixelshuffle} operation for enlarging the spatial resolution.  
	
	\textbf{Flow guidance module.} $F$ is composed of stacked invertible flow cells~\cite{nvp,irn}, which can be defined as $F = f_1\circ f_2 \circ \ldots \circ f_K$. As shown in Figure~\ref{fig:FGM} (b), for the $k^{th}$ cell, the input $l_k$ is split into $l_k^1$ and $l_k^2$ along the channel axis, and they undergo the additive affine transformations~\cite{nice}. Here, to enhance the transformation, we employ an augmented version of the identity branch referring to~\cite{nvp}. The forward and inverse transformations can be expressed as Eq.~\ref{eq:5} and Eq.~\ref{eq:6} respectively:
	\begin{equation}\label{eq:5}
		\small
		\begin{split}
			l_{k+1}^1 = l_k^1 + \alpha(l_k^2), \quad
			l_{k+1}^2 = l_k^2 \otimes \mathrm{exp}(\beta(l_{k+1}^1)) + \varphi(l_{k+1}^1).
		\end{split}
	\end{equation}
\vspace{-0.03\linewidth}
	\begin{equation}\label{eq:6}
		\small
		\begin{split}
		l_k^2 = (l_{k+1}^2 - \varphi(l_{k+1}^1))\otimes \mathrm{exp}(-\beta(l_{k+1}^1)), \quad	
		l_k^1 = l_{k+1}^1 - \alpha(l_k^2)
		\end{split}
	\end{equation}
	where $\alpha (\cdot)$, $\beta (\cdot)$ and $\varphi (\cdot)$ can be arbitrary transformation functions. Here we use Dense Block~\cite{esrgan} for all three functions. Due to the invertibility, $F$ can perform lossless transformation between the downscaled representation and the visually pleasing LR image.
	
	As shown in Figure~\ref{fig:FGRN} (a), during training, the original HR image is firstly downscaled by $G_d$ to the 3-channel representation $x$. Then $x$ will be processed by $G_u$ and $F$ respectively. $G_u$ aims to reconstruct a visually pleasing HR image through $\hat{x}$. While $F$ tries to transform it forward into the visually plausible LR image. During the test (as shown in Figure~\ref{fig:FGRN} (b)), the HR image is downscaled by $G_d$ to the upscaling-optimal representation and then transformed by $F$ to the downscaled LR image, which is the downscaling process. For upscaling, the given LR image is inversely transformed by $F^{-1}$ to the corresponding upscaling-optimal representation, and then it is reconstructed by $G_u$ into the final upscaled image.
	
	\textbf{Quantization}. To convert the values of output LR images from floating-point numbers to 8-bit unsigned integers, we adopt the same quantization module in IRN~\cite{irn}. Although the rounding operation is non-differentiable, the quantization module straightly uses the gradients before rounding to approximate the gradients after rounding. In this way, the quantization module can be inserted into any trainable network. To ensure the transformation of the flow guidance module is lossless and invertible, we also use the same module to convert the downscaled representations to quantized values.
	
	\subsection{Training Objectives}
	Since the joint downscaling-upscaling module and the flow guidance module serve different purposes, we optimize them with different optimization objectives.
	
	\textbf{HR Reconstruction.} Since $G_d$ and $G_u$ are responsible for improving the HR reconstruction performance without considering the visual quality of the downscaled image, we jointly optimize these two modules as follows:
	\begin{equation}\label{eq:7}
		\small
		\begin{split}
			\mathcal{L}_{rec} = \mathbb{E}_{y\sim p_Y}\|y-G_u((G_d(y;\theta_{G_d}));\theta_{G_u})\|_1
		\end{split}
	\end{equation}
	Here we employ the original HR image $y$ as supervision and $L1$ loss as the difference metric.
	
	\textbf{Flow Guidance.} Instead of the commonly used Bicubic LR guidance, we employ flow guidance to supervise the downscaling process. To optimize the invertible flow guidance module $F$, we need to specify each item in Eq~\ref{eq:1}. We assume the prior distribution $p_Z(z)\sim N(y_{bic}, \sigma^2)$.  Here $y_{bic}$ denotes the Gaussian mean of which the value is the bicubically downscaled version of the original HR image, and $\sigma^2$ is the variance. This assumption makes sense because $y_{bic}$ can ensure the visual rationality of the downscaled image, while $\sigma$ increases uncertainty which can alleviate limitations of the Bicubic downscaling method. Thus, the downscaling performance of our FGRN is more likely to outperform Bicubic, which will be experimentally demonstrated in Sec~\ref{sec:diff_loss}. As for the determinant of the Jacobian matrix, the transformation expressed in Eq~\ref{eq:5} can make $\frac{\partial F(x)}{\partial x^T}$ triangular and efficient to calculate the determinant~\cite{nice}. Accordingly, the specific objective of Eq~\ref{eq:1}  for maximizing the log-likelihood can be reformulated as follows:
	\begin{equation}\label{eq:8}
		\small
		\begin{split}
			\mathrm{log}p_X(x) = \mathrm{log}p_Z(F(x)) + \mathrm{log}\vert \mathrm{det}(\frac{\partial F(x)}{\partial x^T}) \vert 
			=-\frac{1}{2\sigma^2} \|F(x)-y_{bic}\|_2 + \eta + \sum_{i=1}^{K}\mathrm{log}\vert \mathrm{det}\frac{\partial f_i(l_{i-1})}{\partial l_{i-1}}\vert
		\end{split}
	\end{equation}
	where $\eta$ is a constant, and we define $l_0 = x$, $l_i = f_i(l_{i-1})$ and $l_K=F(x)$. Thus, we optimize $F$ by modifying the  the negative log-likelihood of Eq~\ref{eq:8}:
	
	\begin{equation}\label{eq:9}
		\small
		\mathcal{L}_{guide} = \mathbb{E}_{x\sim p_X}\|F(x; \theta_F)-y_{bic}\|_1 - \gamma \sum_{i=1}^{K}\mathrm{log}\vert \mathrm{det}\frac{\partial f_i(l_{i-1}; \theta_{f_i})}{\partial l_{i-1}}\vert
	\end{equation}
	where $\gamma$ is a weight coefficient positively related to the variance $\sigma^2$. If $\sigma^2\to 0$, $\mathcal{L}_{guide}$ will approximately degenerate to the LR guidance loss~\cite{task}. Here we set $\gamma=1e-3$.
	
	\subsection{Difference to IRN}
	\begin{figure}[h]
		\centering
		\includegraphics[width=0.75\linewidth]{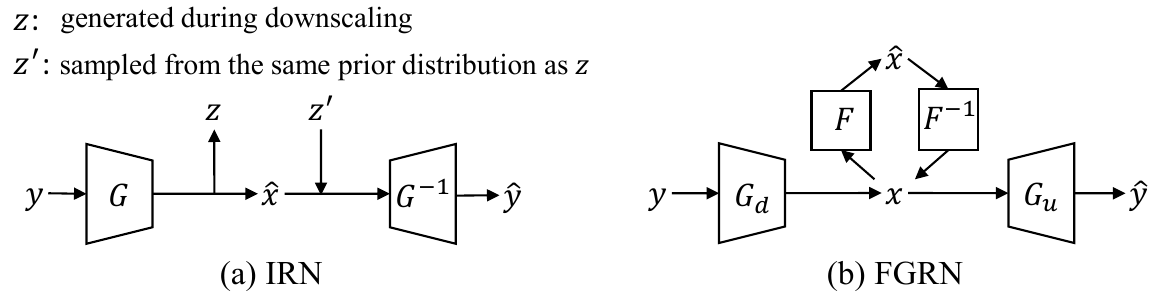}
		\vspace{-0.03\linewidth}
		\caption{The framework of IRN~\cite{irn} (a) and our FGRN (b).}
		\label{f:irn_diff}
		\vspace{-0.02\linewidth}
	\end{figure}
	There are several differences and advantages of FGRN compared to the popular rescaling method IRN~\cite{irn}. 1) As shown in Figure~\ref{f:irn_diff}, IRN models downscaling and upscaling with an invertible network, which is inspired by the reciprocal nature of this pair of rescaling tasks. While the motivation of our FGRN is to find the upscaling-optimal downscaled representation for the encoding-decoding based image rescaling model. 2) IRN relies on an extra latent variable $z$ to help recover the lost information during downscaling. However, We experimentally find that the latent variable fails to formulate the lost information and contributes little to the reconstruction performance. The detailed experiments can be referred to Sec~\ref{sec:info_loss}. While in our work, the lost information could be adaptively recovered by the upscaling module, since the downscaling and upscaling modules are jointly optimized. 3) To keep its invertibility, the downscaling and upscaling modules in IRN must share the same parameters in opposite directions, which may largely limit its representation capability. While our downscaling and upscaling modules are composited of ordinary convolutional layers, which allows us to freely adjust their structures. Besides, the two modules do not share any parameters and potentially have larger representation capability.

	\begin{figure}[h]
		\centering
		\includegraphics[width=0.88\linewidth]{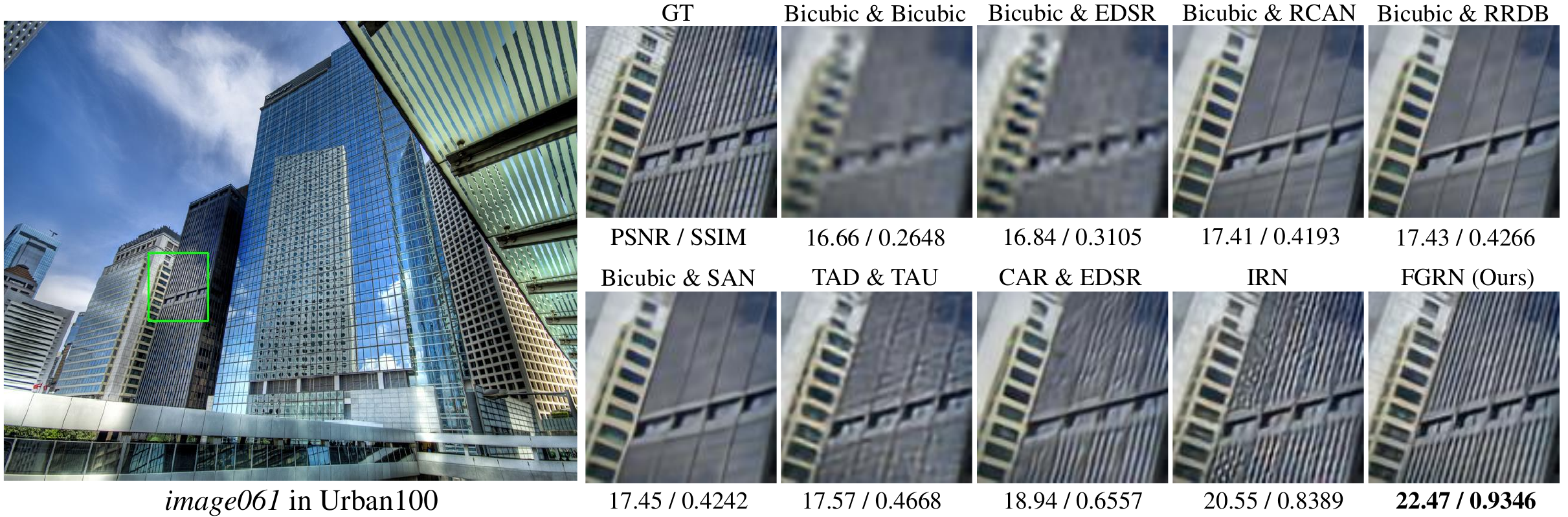}
		\caption{PSNR/SSIM and visual comparison of upscaling the $\times 4$ downscaled image.}
		\label{f:sr}
		\vspace{-0.02\linewidth}
	\end{figure}

	\begin{figure}[h]
		\centering
		\includegraphics[width=0.75\linewidth]{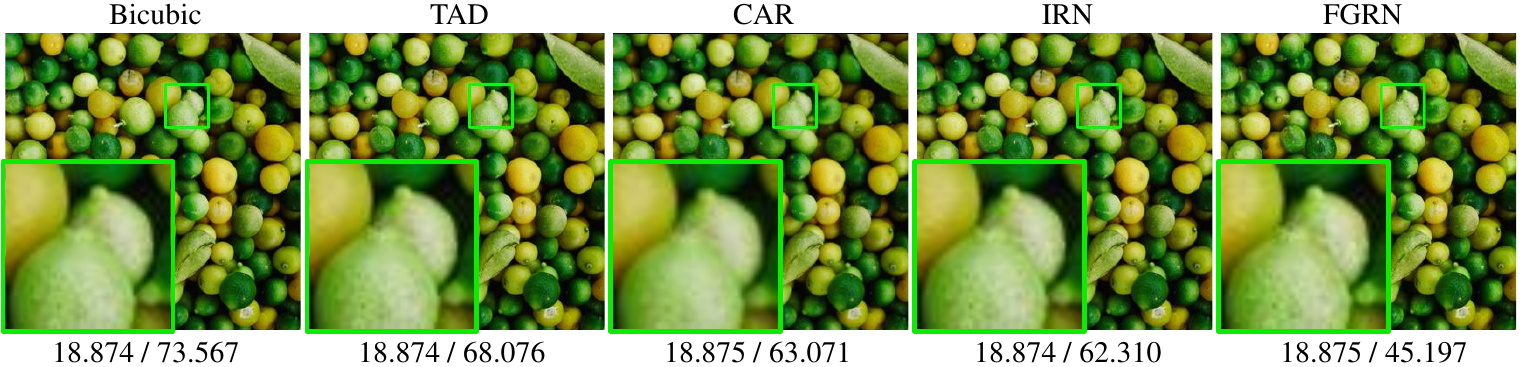}
		\caption{NIQE/PIQE and visual comparison of the $\times 4$ downscaled image.}\label{f:lr}
		\vspace{-0.02\linewidth}
	\end{figure}
	
	\section{Experiments}
	\subsection{Experimental Setup}
	\textbf{Datasets and Metrics.} We employ 3450 high-quality (2K resolution) images as the training data, including 800 images from the training set of DIV2K~\cite{div2k} and 2650 images from Flickr2K~\cite{esrgan}. For the test, we use five standard benchmark datasets: Set5, set14, BSD100, Urban 100, and the validation set of DIV2K. The visual quality of upscaled and downscaled results is evaluated on the Y channel (\ie luminance) of transformed YCbCr space. Since the downscaled results have no corresponding ground truth, they are evaluated with no-reference metrics: Natural Image Quality Evaluator (NIQE)~\cite{niqe} and Perceptual-based Quality Evaluator (PIQE)~\cite{piqe}. Lower NIQE and PIQE indicate higher perceptual quality. While the upscaled results are evaluated with full-reference metrics: PSNR and SSIM~\cite{ssim}.
	
	\noindent\textbf{Implement Details.} For the training data, the batch size is 36 and the size of input HR patches is $192\times 192$. During training, the input images are randomly rotated by $90^\circ$, $180^\circ$ or $270^\circ$ and randomly flipped. We use Adam~\cite{adam} as the optimizer. The initial learning rate is $3\times 10^{-4}$ and decayed by half every $2\times 10^5$ iterations.

	\subsection{Evaluation of Reconstructed HR Images}
	We compare the HR reconstruction performance of FGRN with two kinds of downscaling and upscaling methods: \textit{Type1:} the downscaling module is fixed (\ie Bicubic interpolation) and the upscaling module is the SotA SR method~\cite{srcnn, edsr, rdn, rcan, esrgan, san}; \textit{Type2:} the downscaling module is jointly optimized with the upscaling module~\cite{task, cnncr, resampler, irn}.
	
	\begin{table}[h]
		\small
		\centering
		\caption{Quantitative results of different downscaling and upscaling methods for HR image reconstruction. Bold text: best.}\label{sotahr}
		\setlength{\tabcolsep}{1mm}{
			\resizebox{\textwidth}{!}{
				\begin{tabular}{l|c|c|c|c|c|c|c|c}
					\hline
					& \multirow{2}{*}{Downscaling \& Upscaling}           & \multirow{2}{*}{Scale} & \multirow{2}{*}{Param} & Set5           & Set14          & BSD100         & Urban100       & DIV2K          \\
					&                                                     &                        &                        & PSNR$\uparrow$ / SSIM$\uparrow$      &PSNR$\uparrow$ / SSIM$\uparrow$          &PSNR$\uparrow$ / SSIM$\uparrow$           &PSNR$\uparrow$ / SSIM$\uparrow$         &PSNR$\uparrow$ / SSIM$\uparrow$         \\ \hline \hline
					\multirow{6}{*}{Type1} & Bicubic \& Bicubic                                  &  \multirow{6}{*}{$\times2$}             & -                      & 33.66 / 0.9299 & 30.24 / 0.8688 & 29.56 / 0.8431 & 26.88 / 0.8403 & 31.01 / 0.9393 \\ 
					& Bicubic \& SRCNN~\cite{srcnn} &           & 57.3K                  & 36.66 / 0.9542 & 32.45 / 0.9067 & 31.36 / 0.8879 & 29.50 / 0.8946 & –              \\ 
					& Bicubic \& EDSR~\cite{edsr}   &             & 40.7M                  & 38.20 / 0.9606 & 34.02 / 0.9204 & 32.37 / 0.9018 & 33.10 / 0.9363 & 35.12 / 0.9699 \\ 
					& Bicubic \& RDN~\cite{rdn}     &           & 22.1M                  & 38.24 / 0.9614 & 34.01 / 0.9212 & 32.34 / 0.9017 & 32.89 / 0.9353 & –              \\ 
					& Bicubic \& RCAN~\cite{rcan}  &           & 15.4M                  & 38.27 / 0.9614 & 34.12 / 0.9216 & 32.41 / 0.9027 & 33.34 / 0.9384 & –              \\
					& Bicubic \& SAN~\cite{san}     &            & 15.7M                  & 38.31 / 0.9620 & 34.07 / 0.9213 & 32.42 / 0.9028 & 33.10 / 0.9370 & –              \\ \hline
					\multirow{5}{*}{Type2} & TAD \& TAU~\cite{task}        &     \multirow{5}{*}{$\times2$}          & –                      & 38.46 / –      & 35.52 / –      & 36.68 / –      & 35.03 / –      & 39.01 / –      \\
					& CNN-CR \& CNN-SR~\cite{cnncr}    &            & –                      & 38.88 / –      & 35.40 / –      & 33.92 / –      & 33.68 / –      & –              \\ 
					& CAR \& EDSR~\cite{resampler}       &             & 51.1M                  & 38.94 / 0.9658 & 35.61 / 0.9404 & 33.83 / 0.9262 & 35.24 / 0.9572 & 38.26 / 0.9599 \\  
					& IRN~\cite{irn}               &           & 1.66M                  & 43.99 / 0.9871 & 40.79 / 0.9778 & 41.32 / 0.9876 & 39.92 / 0.9865 & 44.32 / 0.9908 \\ 
					& FGRN (Ours)                                         &             &         1.33M               &    \textbf{45.07 / 0.9896}      &    \textbf{42.28 / 0.9834}       &   \textbf{42.87 / 0.9902}   &   \textbf{ 41.53 / 0.9890}       & \textbf{45.00 / 0.9910}             \\ \hline \hline
					\multirow{7}{*}{Type1} & Bicubic \& Bicubic                                  &  \multirow{7}{*}{$\times4$}            & -                      & 28.42 / 0.8104 & 26.00 / 0.7027 & 25.96 / 0.6675 & 23.14 / 0.6577 & 26.66 / 0.8521 \\
					& Bicubic \& SRCNN~\cite{srcnn}        &          & 57.3K                  & 30.48 / 0.8628 & 27.50 / 0.7513 & 26.90 / 0.7101 & 24.52 / 0.7221 & –              \\ 
					& Bicubic \& EDSR~\cite{edsr}  &             & 43.1M                  & 32.62 / 0.8984 & 28.94 / 0.7901 & 27.79 / 0.7437 & 26.86 / 0.8080 & 29.38 / 0.9032 \\
					& Bicubic \& RDN~\cite{rdn}    &              & 22.3M                  & 32.47 / 0.8990 & 28.81 / 0.7871 & 27.72 / 0.7419 & 26.61 / 0.8028 & –              \\
					& Bicubic \& RCAN~\cite{rcan}   &            & 15.6M                  & 32.63 / 0.9002 & 28.87 / 0.7889 & 27.77 / 0.7436 & 26.82 / 0.8087 & 30.77 / 0.8460 \\  
					& Bicubic \& RRDB~\cite{esrgan} &          & 16.3M                  & 32.74 / 0.9012 & 29.00 / 0.7915 & 27.84 / 0.7455 & 27.03 / 0.8152 & 30.92 / 0.8486 \\ 
					& Bicubic \& SAN~\cite{san}     &            & 15.7M                  & 32.64 / 0.9003 & 28.92 / 0.7888 & 27.78 / 0.7436 & 26.79 / 0.8068 & –              \\ \hline
					\multirow{4}{*}{Type2} & TAD \& TAU~\cite{task}       &  \multirow{4}{*}{$\times4$}           & –                      & 31.81 / –      & 28.63 / –      & 28.51 / –      & 26.63 / –      & 31.16 / –      \\ 
					& CAR \& EDSR~\cite{resampler}       &             & 52.8M                  & 33.88 / 0.9174 & 30.31 / 0.8382 & 29.15 / 0.8001 & 29.28 / 0.8711 & 32.82 / 0.8837 \\ 
					& IRN~\cite{irn}               &             & 4.35M                  & $\mathbf{36.19}$ / 0.9451 & 32.67 / 0.9015 & 31.64 / 0.8826 & 31.41 / 0.9157 & 35.07 / 0.9318 \\ 
					& FGRN (Ours)                                         &             & 3.35M                  & 36.16 / $\mathbf{0.9460}$ &  \textbf{32.99 / 0.9089} &  \textbf{31.83 / 0.8898} &  \textbf{31.47 / 0.9197} & \textbf{35.14 / 0.9341} \\ \hline
				\end{tabular}
		}}
		\vspace{-0.01\linewidth}
	\end{table}	
	
	\noindent\textbf{Quantitative Results}\quad We evaluate the average PSNR and SSIM on five benchmark datasets. Besides, we also provide a comparison between the number of parameters.  As shown in Table~\ref{sotahr}, compared with methods in Type1, the jointly trained methods in Type2 can obtain downscaled images that are more conducive to improving the reconstruction performance. However, the direct LR guidance still restricts the reconstruction performance of the jointly trained methods. In Type2, IRN used to be a strong baseline for image rescaling models, but FGRN can outperform it by a large margin (\eg PSNR:$+1.0\sim1.5dB$ for $\times 2$). Meanwhile, FGRN consumes fewer parameters than other methods. It indicates that the invertible flow guidance can help relax restrictions on the downscaling module and efficiently boost the reconstruction performance.
	
	\noindent\textbf{Qualitative Results} \quad In Figure~\ref{f:sr}, we intuitively present visual comparisons of these methods on upscaling the $4\times$ downscaled image. Methods that the downscaling and upscaling modules are separately trained tend to produce undesirable artifacts and distortion. Even the jointly trained methods can not avoid this problem. In contrast, our FGRN reconstructs the texture clearer and more visually pleasing.

	\subsection{Evaluation of Downscaled LR Images}
	We also compare the downscaling performance of FGRN with other downscaling methods. Since the downscaled results have no corresponding ground truth images, we employ the no-reference NIQE and PIQE as assessment metrics. As shown in Tabel~\ref{sotalr}, our downscaled images have comparable or even better quality than other methods. Besides, Figure~\ref{f:lr} also demonstrates the pleasant visual effects of our method. It indicates that FGRN does not sacrifice the quality of downscaled LR images for the prominent reconstruction performance. When the downscaling factor is larger, the advantage of our method is more obvious. 
	
		\begin{table}[h]
	\vspace{-0.00\linewidth}
	\centering
	\caption{Quantitative results of different downscaling methods. Bold text: best.}\label{sotalr}
	\setlength{\tabcolsep}{3mm}{
		\resizebox{\textwidth}{!}{
			\begin{tabular}{c|c|c|c|c|c|c}
				\hline
				\multirow{2}{*}{Downscaling Method} & \multirow{2}{*}{Scale}      & Set5              & Set14             & BSD100            & Urban100          & DIV2K            \\
				&                             & NIQE$\downarrow$ / PIQE$\downarrow$       & NIQE$\downarrow$ / PIQE$\downarrow$       & NIQE$\downarrow$ / PIQE$\downarrow$      & NIQE$\downarrow$ / PIQE$\downarrow$       & NIQE$\downarrow$ / PIQE$\downarrow$     \\ \hline \hline
				Biucbic                             & \multirow{5}{*}{$\times 2$} & $16.027$ / $38.561$ & $8.519$ / $39.802$  & $5.282$ / $32.494$  & $5.708$ / $42.788$  & ${2.944}$ / $34.618$ \\
				TAD~\cite{task}                                 &                             & $16.027$ / $39.861$ & $\mathbf{8.469}$ / $39.356$  & $5.273$ / $32.180$         & ${5.700}$ / $42.248$  & $2.947$ / $34.357$   \\
				CAR~\cite{resampler}                                 &                             & $16.100$ / $48.496$ & $8.943$ / $41.016$  & $5.859$ / $30.597$  & $6.401$ / $44.446$  & $3.578$ / $35.780$ \\
				IRN~\cite{irn}                                &                             & $\mathbf{16.017}$ / $\mathbf{34.623}$ & $8.558$ / $\mathbf{37.001}$  & $5.818$ / $\mathbf{30.046}$  & $6.143$ / $\mathbf{38.161}$  & $3.391$ / $\mathbf{30.244}$ \\
				FGRN (Ours)                         &                             & $16.103$ / ${40.762}$ & $8.597$ / ${38.396}$  & $\mathbf{5.088}$ / $31.640$  & $\mathbf{5.630}$ / ${42.139}$  & $\mathbf{2.928}$ / $34.784$ \\ \hline \hline
				Biucbic                             & \multirow{5}{*}{$\times 4$} & $18.876$ / $51.101$ & $17.932$ / $45.335$ & $18.878$ / $41.527$ & $17.378$ / $46.609$ & $3.983$ / $38.161$ \\
				TAD~\cite{task}                                 &                             & $18.876$ / $49.691$ & $17.936$ / $45.016$ & $18.878$ / $41.190$ & $17.320$ / $46.360$ & $\mathbf{3.964}$ / $37.784$               \\
				CAR~\cite{resampler}                                 &                             & $\mathbf{18.873}$ / $59.884$ & $17.922$ / $45.479$ & $18.878$ / $41.431$ & $20.284$ / $48.508$ & $5.499$ / $37.821$ \\
				IRN~\cite{irn}                                 &                             & $18.876$ / $45.112$ & $\mathbf{17.909}$ / $40.213$ & $18.879$ / $35.713$ & $17.755$ / $39.769$ & $4.292$ / $30.233$ \\
				FGRN (Ours)                         &                             & $18.875$ / $\mathbf{45.075}$ & $17.918$ / $\mathbf{38.727}$ & $\mathbf{18.878}$ / $\mathbf{35.218}$ & $\mathbf{17.276}$ / $\mathbf{37.307  }$ & $4.055$ / $\mathbf{29.256}$ \\ \hline
			\end{tabular}
	}}
	\vspace{-0.02\linewidth}
\end{table}

	\subsection{Study on the Downscaling Guidance} \label{sec:diff_loss}
\begin{figure}[h]
		\centering
		\includegraphics[width=0.95\linewidth]{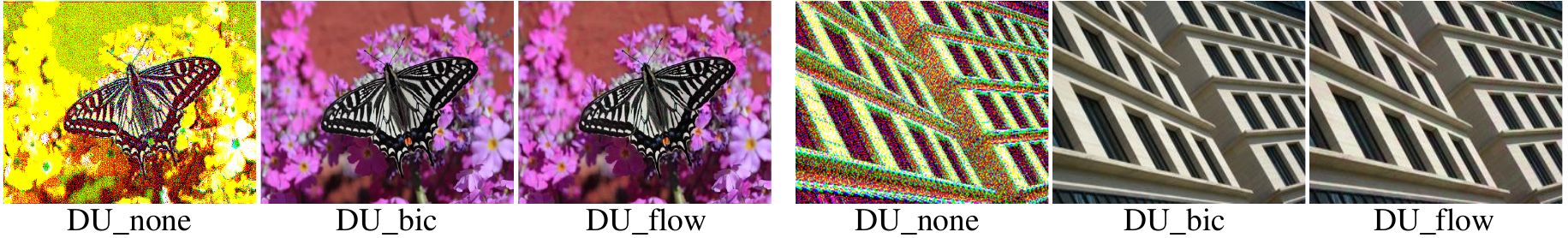}
		\vspace{-0.01\linewidth}
		\caption{Visual results of methods with different downscaling guidance.}
		\label{f:diff_loss}
		\vspace{-0.04\linewidth}
	\end{figure}

	We experimentally investigate the influence of different downscaling LR guidances on the HR reconstruction performance. The baseline model is the pure encoding-decoding model without any downscaling guidance, denoted as \textit{DU\_none}. \textit{DU\_bic} uses the bicubically downscaled image as the guidance, \ie bicubic LR guidance, while \textit{DU\_flow} use our flow guidance as the downscaling guidance to supervise the downscaling process. The downscaling and reconstruction performance are shown in Figure~\ref{f:diff_loss} and Table~\ref{diff_loss} respectively.
	
	\textit{DU\_none} only needs to focus on enhancing the HR image quality from the upscaling module, which can maximize the reconstruction performance. The reconstruction performance of DU\_none in Table~\ref{diff_loss} can be seen as the upper limit. However, without the restriction of downscaling guidance, its downscaling performance is unsatisfactory as illustrated in Figure~\ref{f:diff_loss}. The bicubic LR guidance in \textit{DU\_bic} can largely improve the quality of the downscaled LR image, but the reconstruction performance is compromised because the visually favorable downscaled image is not optimal for the upscaling process. As for the \textit{DU\_flow}, both excellent downscaling performance and HR reconstruction performance can be achieved. It indicates the superiority of our flow guidance that can decouple the mutual interference of the downscaling and upscaling, thus maximizing the capabilities of both processes.
		
	\begin{table}[h]
		\centering
		\caption{Reconstruction results of training the joint downscling-upscaling model with differen downscaling guidance.}\label{diff_loss}
		\setlength{\tabcolsep}{3mm}{
			\resizebox{0.95\textwidth}{!}{
				\begin{tabular}{c|c|c|c|c|c|c}
					\hline
					& \multirow{2}{*}{Scale}      & Set5           & Set14          & BSD100         & Urban100       & DIV2K          \\
					&                             & PSNR$\uparrow$ / SSIM$\uparrow$   &  PSNR$\uparrow$ / SSIM$\uparrow$    & PSNR$\uparrow$ / SSIM$\uparrow$    &  PSNR$\uparrow$ / SSIM$\uparrow$    &  PSNR$\uparrow$ / SSIM$\uparrow$   \\ \hline \hline
					DU\_none   & \multirow{3}{*}{$\times 4$} & 36.91 / 0.9538 & 33.79 / 0.9215 & 32.41 / 0.9045 & 32.07 / 0.9307 & 35.35 / 0.9338 \\
					DU\_bic &                             & 34.31 / 0.9258 & 30.95 / 0.8645 & 29.99 / 0.8365 & 29.38 / 0.8823 & 33.14 / 0.9020 \\
					DU\_flow   &                              &36.16 / 0.9460 &  32.99 / 0.9089 &  31.83 / 0.8898 &  31.47 / 0.9197 & 35.14 / 0.9341 \\ \hline
				\end{tabular}
		}}
		\vspace{-0.02\linewidth}
	\end{table}
	
	\subsection{Study on the Information Loss of IRN} \label{sec:info_loss}
	IRN claims that the high-frequency information lost during downscaling could be embedded into a latent variable $z$ obeying a prior distribution, which could compensate for information during upscaling. In this section, we experimentally investigate the role of the latent variable for IRN. We fix the latent variable as 0 (denoted as $z_0$) and retrain IRN. In this case, $z_0$ becomes a part of the parameters of the upscaling module. As shown in Table~\ref{fix_z}, the retrained IRN with $z_0$ can still achieve the same results as the original one with $z$ obeying Gaussian distribution. It indicates that the latent variable in IRN could not provide specific high-frequency information for different images to improve the reconstruction performance. The lost information in IRN is actually implicitly learned in the upscaling module.

	\begin{table}[h]
		\centering
		\caption{Reconstruction results of training IRN~\cite{irn} with different latent variable.}\label{fix_z}
		\setlength{\tabcolsep}{3mm}{
			\resizebox{0.95\textwidth}{!}{
				\begin{tabular}{c|c|c|c|c|c|c}
					\hline
					\multirow{2}{*}{} & \multirow{2}{*}{Scale} & Set5           & Set14          & BSD100         & Urban100       & DIV2K                         \\
					&                        & PSNR$\uparrow$ / SSIM $\uparrow$      & PSNR$\uparrow$ / SSIM$\uparrow$      & PSNR$\uparrow$ / SSIM$\uparrow$      & PSNR$\uparrow$ / SSIM$\uparrow$      & \multicolumn{1}{c}{PSNR $\uparrow$ / SSIM$\uparrow$} \\ \hline\hline
					$z$                & \multirow{2}{*}{ $\times$4 }     & 36.19 / 0.9451 & 32.67 / 0.9015 & 31.64 / 0.8826 & 31.41 / 0.9157 & 35.07 / 0.9318                \\
					$z_0$                &                        & 36.21 / 0.9455 & 32.70 / 0.9013 & 31.62 / 0.8820 & 31.36 / 0.9147 & 35.16 / 0.9316                \\ \hline
				\end{tabular}
		}}
		\vspace{-0.03\linewidth}
	\end{table}

	\section{Conclusion}
	In this paper, we argue that in image rescaling, commonly used bicubic LR guidance will limit the reconstruction performance of the subsequent upscaling task. Towards this issue, we propose a flow guidance image rescaling network (FGRN) that can eliminate the restriction of the bicubic LR guidance. In our method, the invertible flow guidance module (FGM) can make a lossless transformation between the downscaled representation and the visually pleasing LR image, which can not only guarantee the quality of the downscaled image but also remove the restrictions on the downscaled representation. Thus, the downscaling and upscaling can be jointly optimized in an end-to-end manner without considering the visual quality of the downscaled image.  In this way, the downscaled representation can become upscaling-optimal, which makes the reconstruction capability of the upscaling module approach the upper limit. Extensive experiments demonstrate that our model significantly improves both downscaling and HR reconstruction performance while being light-weighted.
	\\

	\noindent\textbf{Acknowledgments}\quad This work was supported by the National Key R\&D Program of China (2019YFB1406200). It was also the research achievement of the Key Laboratory of Digital Rights Services.


\begin{thebibliography}{36}
	\providecommand{\natexlab}[1]{#1}
	\providecommand{\url}[1]{\texttt{#1}}
	\expandafter\ifx\csname urlstyle\endcsname\relax
	\providecommand{\doi}[1]{doi: #1}\else
	\providecommand{\doi}{doi: \begingroup \urlstyle{rm}\Url}\fi
	
	\bibitem[Agustsson and Timofte(2017)]{div2k}
	Eirikur Agustsson and Radu Timofte.
	\newblock Ntire 2017 challenge on single image super-resolution: Dataset and
	study.
	\newblock In \emph{Proceedings of the IEEE Conference on Computer Vision and
		Pattern Recognition Workshops}, pages 126--135, 2017.
	
	\bibitem[Chen et~al.(2020)Chen, Gong, Wang, Li, and Wong]{facesr}
	Chaofeng Chen, Dihong Gong, Hao Wang, Zhifeng Li, and Kwan-Yee~K Wong.
	\newblock Learning spatial attention for face super-resolution.
	\newblock \emph{IEEE Transactions on Image Processing}, 30:\penalty0
	1219--1231, 2020.
	
	\bibitem[Dai et~al.(2019)Dai, Cai, Zhang, Xia, and Zhang]{san}
	Tao Dai, Jianrui Cai, Yongbing Zhang, Shu-Tao Xia, and Lei Zhang.
	\newblock Second-order attention network for single image super-resolution.
	\newblock In \emph{Proceedings of the IEEE/CVF Conference on Computer Vision
		and Pattern Recognition}, pages 11065--11074, 2019.
	
	\bibitem[Dinh et~al.(2014)Dinh, Krueger, and Bengio]{nice}
	Laurent Dinh, David Krueger, and Yoshua Bengio.
	\newblock Nice: Non-linear independent components estimation.
	\newblock \emph{arXiv preprint arXiv:1410.8516}, 2014.
	
	\bibitem[Dinh et~al.(2016)Dinh, Sohl-Dickstein, and Bengio]{nvp}
	Laurent Dinh, Jascha Sohl-Dickstein, and Samy Bengio.
	\newblock Density estimation using real nvp.
	\newblock \emph{arXiv preprint arXiv:1605.08803}, 2016.
	
	\bibitem[Dong et~al.(2014)Dong, Loy, He, and Tang]{srcnn}
	Chao Dong, Chen~Change Loy, Kaiming He, and Xiaoou Tang.
	\newblock Learning a deep convolutional network for image super-resolution.
	\newblock In \emph{European conference on computer vision}, pages 184--199.
	Springer, 2014.
	
	\bibitem[Glasner et~al.(2009)Glasner, Bagon, and Irani]{glasner}
	Daniel Glasner, Shai Bagon, and Michal Irani.
	\newblock Super-resolution from a single image.
	\newblock In \emph{2009 IEEE 12th international conference on computer vision},
	pages 349--356. IEEE, 2009.
	
	\bibitem[Grover et~al.(2018)Grover, Dhar, and Ermon]{flowgan}
	Aditya Grover, Manik Dhar, and Stefano Ermon.
	\newblock Flow-gan: Combining maximum likelihood and adversarial learning in
	generative models.
	\newblock In \emph{Proceedings of the AAAI Conference on Artificial
		Intelligence}, volume~32, 2018.
	
	\bibitem[Han et~al.(2021)Han, Huang, Song, Wang, and Tan]{reidsr}
	Ke~Han, Yan Huang, Chunfeng Song, Liang Wang, and Tieniu Tan.
	\newblock Adaptive super-resolution for person re-identification with
	low-resolution images.
	\newblock \emph{Pattern Recognition}, 114:\penalty0 107682, 2021.
	
	\bibitem[Hou et~al.(2017)Hou, Duan, and Qiu]{hou}
	Xianxu Hou, Jiang Duan, and Guoping Qiu.
	\newblock Deep feature consistent deep image transformations: Downscaling,
	decolorization and hdr tone mapping.
	\newblock \emph{arXiv preprint arXiv:1707.09482}, 2017.
	
	\bibitem[Kim et~al.(2018{\natexlab{a}})Kim, Choi, Lim, and Lee]{task}
	Heewon Kim, Myungsub Choi, Bee Lim, and Kyoung~Mu Lee.
	\newblock Task-aware image downscaling.
	\newblock In \emph{Proceedings of the European Conference on Computer Vision
		(ECCV)}, pages 399--414, 2018{\natexlab{a}}.
	
	\bibitem[Kim et~al.(2016)Kim, Lee, and Lee]{accuratesr}
	Jiwon Kim, Jung~Kwon Lee, and Kyoung~Mu Lee.
	\newblock Accurate image super-resolution using very deep convolutional
	networks.
	\newblock In \emph{Proceedings of the IEEE conference on computer vision and
		pattern recognition}, pages 1646--1654, 2016.
	
	\bibitem[Kim et~al.(2018{\natexlab{b}})Kim, Lee, Song, Kim, and
	Yoon]{flowavenet}
	Sungwon Kim, Sang-Gil Lee, Jongyoon Song, Jaehyeon Kim, and Sungroh Yoon.
	\newblock Flowavenet: A generative flow for raw audio.
	\newblock \emph{arXiv preprint arXiv:1811.02155}, 2018{\natexlab{b}}.
	
	\bibitem[Kingma and Ba(2014)]{adam}
	Diederik~P Kingma and Jimmy Ba.
	\newblock Adam: A method for stochastic optimization.
	\newblock \emph{arXiv preprint arXiv:1412.6980}, 2014.
	
	\bibitem[Kingma and Dhariwal(2018)]{glow}
	Diederik~P Kingma and Prafulla Dhariwal.
	\newblock Glow: Generative flow with invertible 1x1 convolutions.
	\newblock \emph{arXiv preprint arXiv:1807.03039}, 2018.
	
	\bibitem[Kopf et~al.(2013)Kopf, Shamir, and Peers]{kopf}
	Johannes Kopf, Ariel Shamir, and Pieter Peers.
	\newblock Content-adaptive image downscaling.
	\newblock \emph{ACM Transactions on Graphics (TOG)}, 32\penalty0 (6):\penalty0
	1--8, 2013.
	
	\bibitem[Ledig et~al.(2017)Ledig, Theis, Husz{\'a}r, Caballero, Cunningham,
	Acosta, Aitken, Tejani, Totz, Wang, et~al.]{srgan}
	Christian Ledig, Lucas Theis, Ferenc Husz{\'a}r, Jose Caballero, Andrew
	Cunningham, Alejandro Acosta, Andrew Aitken, Alykhan Tejani, Johannes Totz,
	Zehan Wang, et~al.
	\newblock Photo-realistic single image super-resolution using a generative
	adversarial network.
	\newblock In \emph{Proceedings of the IEEE conference on computer vision and
		pattern recognition}, pages 4681--4690, 2017.
	
	\bibitem[Li et~al.(2018)Li, Liu, Li, Li, Li, and Wu]{cnncr}
	Yue Li, Dong Liu, Houqiang Li, Li~Li, Zhu Li, and Feng Wu.
	\newblock Learning a convolutional neural network for image compact-resolution.
	\newblock \emph{IEEE Transactions on Image Processing}, 28\penalty0
	(3):\penalty0 1092--1107, 2018.
	
	\bibitem[Lim et~al.(2017)Lim, Son, Kim, Nah, and Mu~Lee]{edsr}
	Bee Lim, Sanghyun Son, Heewon Kim, Seungjun Nah, and Kyoung Mu~Lee.
	\newblock Enhanced deep residual networks for single image super-resolution.
	\newblock In \emph{Proceedings of the IEEE conference on computer vision and
		pattern recognition workshops}, pages 136--144, 2017.
	
	\bibitem[Liu et~al.(2019)Liu, Liu, Gong, Wang, and Li]{liuconditional}
	Rui Liu, Yu~Liu, Xinyu Gong, Xiaogang Wang, and Hongsheng Li.
	\newblock Conditional adversarial generative flow for controllable image
	synthesis.
	\newblock In \emph{Proceedings of the IEEE/CVF Conference on Computer Vision
		and Pattern Recognition}, pages 7992--8001, 2019.
	
	\bibitem[Luo et~al.(2020)Luo, Huang, Li, Wang, and Tan]{dan}
	Zhengxiong Luo, Yan Huang, Shang Li, Liang Wang, and Tieniu Tan.
	\newblock Unfolding the alternating optimization for blind super resolution.
	\newblock \emph{arXiv preprint arXiv:2010.02631}, 2020.
	
	\bibitem[Michaeli and Irani(2013)]{nonpara}
	Tomer Michaeli and Michal Irani.
	\newblock Nonparametric blind super-resolution.
	\newblock In \emph{Proceedings of the IEEE International Conference on Computer
		Vision}, pages 945--952, 2013.
	
	\bibitem[Mitchell and Netravali(1988)]{bicubic}
	Don~P Mitchell and Arun~N Netravali.
	\newblock Reconstruction filters in computer-graphics.
	\newblock \emph{ACM Siggraph Computer Graphics}, 22\penalty0 (4):\penalty0
	221--228, 1988.
	
	\bibitem[Mittal et~al.(2012)Mittal, Soundararajan, and Bovik]{niqe}
	Anish Mittal, Rajiv Soundararajan, and Alan~C Bovik.
	\newblock Making a “completely blind” image quality analyzer.
	\newblock \emph{IEEE Signal processing letters}, 20\penalty0 (3):\penalty0
	209--212, 2012.
	
	\bibitem[Oeztireli and Gross(2015)]{og}
	A~Cengiz Oeztireli and Markus Gross.
	\newblock Perceptually based downscaling of images.
	\newblock \emph{ACM Transactions on Graphics (TOG)}, 34\penalty0 (4):\penalty0
	1--10, 2015.
	
	\bibitem[Pumarola et~al.(2020)Pumarola, Popov, Moreno-Noguer, and
	Ferrari]{cflow}
	Albert Pumarola, Stefan Popov, Francesc Moreno-Noguer, and Vittorio Ferrari.
	\newblock C-flow: Conditional generative flow models for images and 3d point
	clouds.
	\newblock In \emph{Proceedings of the IEEE/CVF Conference on Computer Vision
		and Pattern Recognition}, pages 7949--7958, 2020.
	
	\bibitem[Shannon(1949)]{nyquist}
	Claude~Elwood Shannon.
	\newblock Communication in the presence of noise.
	\newblock \emph{Proceedings of the IRE}, 37\penalty0 (1):\penalty0 10--21,
	1949.
	
	\bibitem[Shi et~al.(2016)Shi, Caballero, Husz{\'a}r, Totz, Aitken, Bishop,
	Rueckert, and Wang]{pixelshuffle}
	Wenzhe Shi, Jose Caballero, Ferenc Husz{\'a}r, Johannes Totz, Andrew~P Aitken,
	Rob Bishop, Daniel Rueckert, and Zehan Wang.
	\newblock Real-time single image and video super-resolution using an efficient
	sub-pixel convolutional neural network.
	\newblock In \emph{Proceedings of the IEEE conference on computer vision and
		pattern recognition}, pages 1874--1883, 2016.
	
	\bibitem[Sun and Chen(2020)]{resampler}
	Wanjie Sun and Zhenzhong Chen.
	\newblock Learned image downscaling for upscaling using content adaptive
	resampler.
	\newblock \emph{IEEE Transactions on Image Processing}, 29:\penalty0
	4027--4040, 2020.
	
	\bibitem[Venkatanath et~al.(2015)Venkatanath, Praneeth, Bh, Channappayya, and
	Medasani]{piqe}
	N~Venkatanath, D~Praneeth, Maruthi~Chandrasekhar Bh, Sumohana~S Channappayya,
	and Swarup~S Medasani.
	\newblock Blind image quality evaluation using perception based features.
	\newblock In \emph{2015 Twenty First National Conference on Communications
		(NCC)}, pages 1--6. IEEE, 2015.
	
	\bibitem[Wang et~al.(2018)Wang, Yu, Wu, Gu, Liu, Dong, Qiao, and
	Change~Loy]{esrgan}
	Xintao Wang, Ke~Yu, Shixiang Wu, Jinjin Gu, Yihao Liu, Chao Dong, Yu~Qiao, and
	Chen Change~Loy.
	\newblock Esrgan: Enhanced super-resolution generative adversarial networks.
	\newblock In \emph{Proceedings of the European Conference on Computer Vision
		(ECCV) Workshops}, pages 0--0, 2018.
	
	\bibitem[Wang et~al.(2004)Wang, Bovik, Sheikh, and Simoncelli]{ssim}
	Zhou Wang, Alan~C Bovik, Hamid~R Sheikh, and Eero~P Simoncelli.
	\newblock Image quality assessment: from error visibility to structural
	similarity.
	\newblock \emph{IEEE transactions on image processing}, 13\penalty0
	(4):\penalty0 600--612, 2004.
	
	\bibitem[Xiao et~al.(2020)Xiao, Zheng, Liu, Wang, He, Ke, Bian, Lin, and
	Liu]{irn}
	Mingqing Xiao, Shuxin Zheng, Chang Liu, Yaolong Wang, Di~He, Guolin Ke, Jiang
	Bian, Zhouchen Lin, and Tie-Yan Liu.
	\newblock Invertible image rescaling.
	\newblock In \emph{European Conference on Computer Vision}, pages 126--144.
	Springer, 2020.
	
	\bibitem[Yang et~al.(2019)Yang, Huang, Hao, Liu, Belongie, and
	Hariharan]{pointflow}
	Guandao Yang, Xun Huang, Zekun Hao, Ming-Yu Liu, Serge Belongie, and Bharath
	Hariharan.
	\newblock Pointflow: 3d point cloud generation with continuous normalizing
	flows.
	\newblock In \emph{Proceedings of the IEEE/CVF International Conference on
		Computer Vision}, pages 4541--4550, 2019.
	
	\bibitem[Zhang et~al.(2018{\natexlab{a}})Zhang, Li, Li, Wang, Zhong, and
	Fu]{rcan}
	Yulun Zhang, Kunpeng Li, Kai Li, Lichen Wang, Bineng Zhong, and Yun Fu.
	\newblock Image super-resolution using very deep residual channel attention
	networks.
	\newblock In \emph{Proceedings of the European conference on computer vision
		(ECCV)}, pages 286--301, 2018{\natexlab{a}}.
	
	\bibitem[Zhang et~al.(2018{\natexlab{b}})Zhang, Tian, Kong, Zhong, and Fu]{rdn}
	Yulun Zhang, Yapeng Tian, Yu~Kong, Bineng Zhong, and Yun Fu.
	\newblock Residual dense network for image super-resolution.
	\newblock In \emph{Proceedings of the IEEE conference on computer vision and
		pattern recognition}, pages 2472--2481, 2018{\natexlab{b}}.
	
\end{thebibliography}
\end{document}